\documentclass[intlimits,twoside,a4paper]{article}

\usepackage{amsmath,amssymb}
\usepackage{graphicx}
\usepackage[T2A]{fontenc}
\usepackage[cp1251]{inputenc}

\usepackage[eqsecnum]{cmpj2}

\issue{2015}{18}{3}{33703}
\doinumber{10.5488/CMP.18.33703}

\title[Numerical study of the spin-3/2 Ashkin-Teller model]%
{Numerical study of the spin-3/2 Ashkin-Teller model}

\author[R. Boudefla, S. Bekhechi, F. Hontinfinde]%
{R. Boudefla\refaddr{label1}\thanks{raniaboudefla@yahoo.com}\,, S. Bekhechi\refaddr{label1}, F. Hontinfinde\refaddr{label2}
}
\addresses{
\addr{label1} Laboratoire de Physique Tht\'{e}orique (LPT), B.P. 230, Universit\'{e} Abou Bekr Belka\"{\i}d, Tlemcen 13000, Algeria
\addr{label2} D\'{e}partement de Physique (FAST) et Institut des Math\'{e}matiques et de Sciences Physiques (IMSP), Universit\'{e} d'Abomey-Calavy, 01 B.P. 613, Porto-Novo, Benin}

\authorcopyright{R. Boudefla, S. Bekhechi, F. Hontinfinde, 2015}
\date{Received November 8, 2014, in final form May 10, 2015}

\begin{document}

\maketitle

\begin{abstract}
The study of the Ashkin-Teller model (ATM) of spin-3/2 on a hypercubic lattice is undertaken via Monte Carlo simulation. The phase diagrams are displayed and discussed in the physical parameter space.  Rich physical properties are recovered, namely the second order transition and multicritical points. The phase diagrams have been obtained by varying the strength describing the four spin interaction and the single ion potential. This model shows a new high temperature partially ordered phase, called $\langle S\rangle$ and a new Baxter 3/2 ground state which do not exist either in the spin-1/2 ATM or in the spin-1 ATM.
\pacs 75.10.Hk
\end{abstract}

\section{Introduction}

In this work, we will analyze a magnetic model with three spin states known as Ashkin-Teller model \cite{Ash43} which is a superposition of two Ising models with spin variables $\sigma$ and $S$. In every site $i$ of the cubic lattice, two spin variables $\sigma_{i}$ and $S_{i}$ are associated. In each Ising model, there are two spin nearest-neighbors interaction with a strength $K_{2}$ \cite{Kog79}. In addition, different Ising models are coupled by a four-spin interaction with strength $K_{4}$ \cite{Bar13} and on each site there is a single ion potential $D$ \cite{Kog79}. All these interactions are limited to the first nearest neighbors.

Recent researches of this Ising model and its phase diagrams with four spin interaction and some of its applications have been done \cite{Cha14,Hua13,Zhe08,Jia05,Bez01}.

The selenium compound  adsorbed on a nickel surface \cite{Bak85} is a good physical picture for this model. Different methods have been used to understand  the critical behaviour of this model. For the two dimensional case, all of mean-field approximation (MFA) \cite{Zha93,Dit80,Bek99,Bek00} Monte Carlo simulations (MC) \cite{Dit80,Bek99,Bek00,Bad99}, series analysis \cite{Dit72}, exact duality \cite{Weg72}, transfer-matrix finite size scaling calculations \cite{Bak85,Bad99,Bad00}, renormalization group \cite{Ban79,Kno75} and mean field renormalization group approach \cite{Pla86}, yield three different phases: a paramagnetic phase in which neither $\sigma$ nor $S$ nor anything else is ordered $(\langle\sigma \rangle= \langle S \rangle = \langle \sigma S\rangle = 0)$; Baxter phase in which $\sigma$ and $S$ independently order in ferromagnetic fashion, and also $\langle\sigma S\rangle  \neq 0$; a third phase called PO$_{1}$ in which  $\sigma S$ is  ferromagnetically ordered $\langle\sigma S\rangle  \neq 0$ but $ \langle\sigma \rangle= \langle S \rangle =0$.

One of the most interesting and challenging phenomena is the appearance of other new partially ordered phases in the ATM. For example, MFA and MC simulations applied to the three-dimensional case yield first and second-order phase transitions and partially ferromagnetic ordered phase $\langle\sigma \rangle$ ($\langle\sigma \rangle  \neq 0$  and
$\langle S \rangle = \langle\sigma S \rangle = 0$) \cite{Dit80}. By using exact duality transformations and symmetry considerations \cite{Wu74,Bad00}, the anisotropic ATM in $d =2$ also presents partially ordered phases called $\langle\sigma \rangle$ and $\langle S \rangle$  which are connected by symmetry operations to the $\langle\sigma S \rangle $ phase. These results are confirmed in $d= 2$ and $d= 3$ by MFA and MC simulations \cite{Bek99}. The PO$_{2}$ phase defined by ($\langle S \rangle = \langle\sigma \rangle \neq 0$; $\langle\sigma S \rangle = 0$) found in the spin-1 Ashkin-Teller model \cite{Lou97,Bad99}  does not occur in the spin-1/2 Ashkin-Teller model \cite{Bek99}.

Monte Carlo (MC) simulation can be shown as a powerful and successful tool to study critical phenomena \cite{Bek99} at reduced dimensionality ($d= 2$). So, it is of importance to fully understand the phase diagram obtained from this model through a nonperturbative method, such as Monte Carlo technique. The main problem which arises from this method is the existence of statistical errors.

In this paper, we mainly focused on the spin-3/2 Ashkin-Teller model using MC simulations. The paper is organized as follows: in the second section, the investigated model is introduced and the ground state diagram is presented. Section~\ref{sec:Monte-carlo} contains the description  of the methodology  and formalism of the MC  simulations. We  collect our results and discussions in section~\ref{sec:Results}. Finally, the summary and conclusions  are drawn in section~\ref{sec:Conclusion}.

\section{Model and ground state diagram}
\label{sec:2}

The  Hamiltonian of the model can be expressed as:
\begin{equation}
H=-K_{2}\sum_{\langle ij\rangle}\left(\sigma_{i}\sigma_{j}+S_{i}S_{j}\right)-K_{4}\sum_{\langle ij\rangle}\sigma_{i}\sigma_{j}S_{i}S_{j}-D\sum_{\langle i\rangle}\left(S_i^2+\sigma_i^2\right),
\end{equation}
where the spins $\sigma_{i}$ and $S_{i}$ are located on sites of an hypercubic lattice and take both the values $\pm3/2$ and $ \pm1/2$. The first term describes the bilinear interactions between the spins at sites $i$ and $j$, with the interaction parameter $K_{2}$. The second term describes the four-spins interaction with strength $K_{4}$ and on each site there is a single ion potential $D$. All these interactions are restricted to the $z$ nearest neighbours pairs of spins.

In order to calculate the ground state energy, we express the hamiltonian as a sum of contributions of the nearest-neighbour spins. So, the contribution of a pair $S_{1}$, $S_{2}$ and $\sigma_{1}$, $\sigma_{2}$ is:
\begin{equation}
E_{p}=-K_{2}(\sigma_{1}\sigma_{2}+S_{1}S_{2})
-K_{4}\sigma_{1}\sigma_{2}S_{1}S_{2}-D\left(S_1^2+S_2^2+\sigma_1^2+\sigma_2^2\right).
\end{equation}

\begin{figure}[!t]
\centerline{\includegraphics[width=0.6\textwidth]{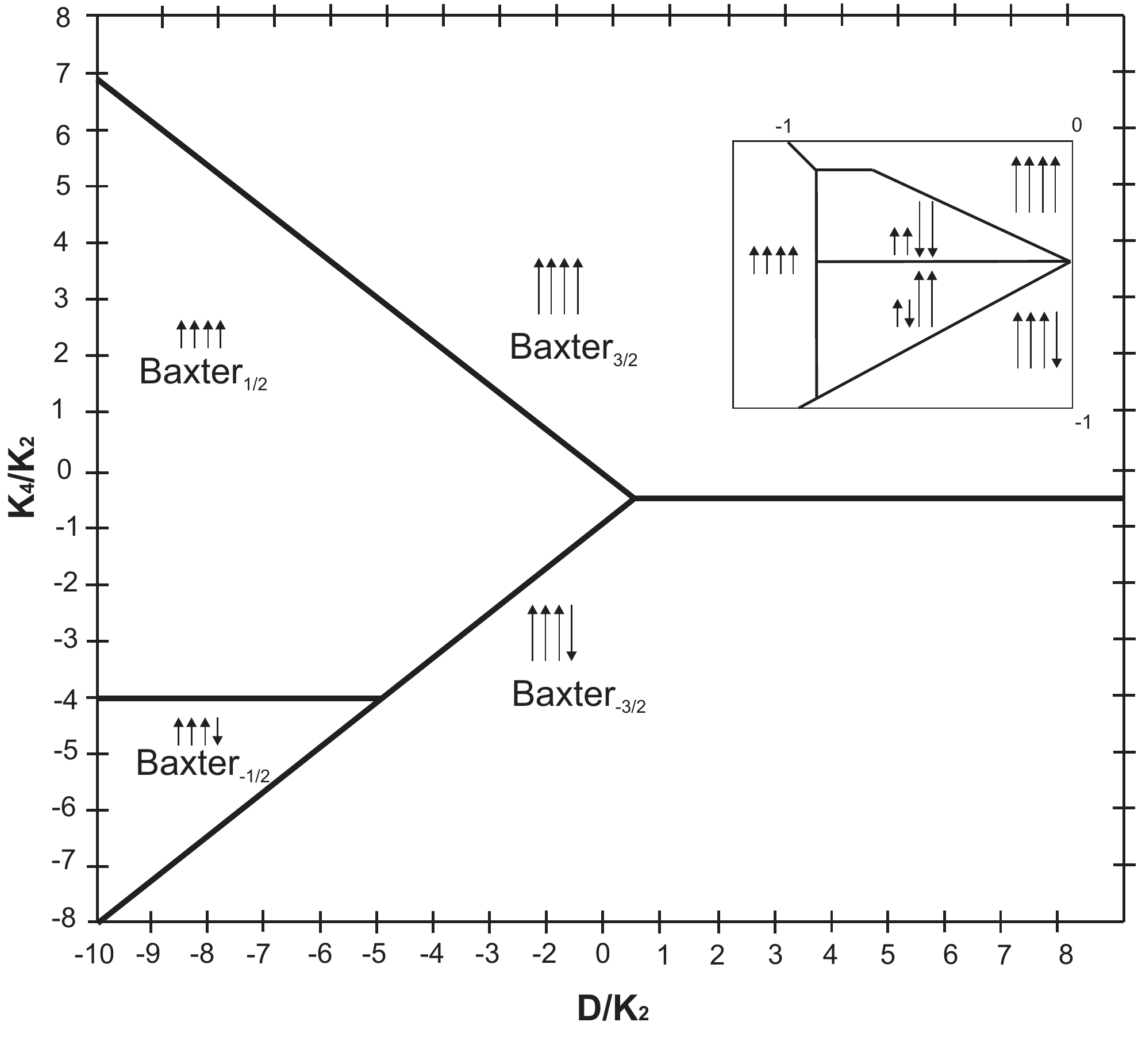}}
\vspace{-3mm}
\caption{Ground state phase diagram.} \label{1}
\end{figure}

By comparing the values of $E_{p}$ for different configurations, we obtain the following structure of phase diagram shown in figure~\ref{1}:
\begin{itemize}
\item[(i)] For $K_{4}/K_{2}<D/K_{2}$: if $K_{4}/K_{2}<-0.4$, the spins $\sigma_{i}$ are parallel while the spins $S_{i}$ are antiparallel. Then we have: $\langle S \rangle_\textrm{F} = \langle\sigma \rangle_\textrm{AF} =\langle\sigma S \rangle_\textrm{F} = 0$ and $\langle S \rangle_\textrm{AF} = \langle\sigma \rangle_\textrm{F} =3/2$ and $\langle\sigma S \rangle_\textrm{AF}  \neq 0$ which characterize the phase called Baxter-3/2, otherwise if $K_{4}/K_{2}>-0.4$ the Baxter-3/2 phase is stable since both spins $\sigma_{i}$ and $S_{i}$ are aligned and equal to 3/2.
\item[(ii)] For $K_{4}/K_{2}>D/K_{2}$: if $K_{4}/K_{2}<-4$, the spins $\sigma_{i}$ are parallel while the spins $S_{i}$ are antiparallel. Then we have: $\langle S \rangle_\textrm{F} = \langle\sigma \rangle_\textrm{AF} =\langle\sigma S \rangle_\textrm{F} = 0$ and $\langle S \rangle_\textrm{AF} = \langle\sigma \rangle_\textrm{F} =1/2$ and $\langle\sigma S \rangle_\textrm{AF}  \neq 0$ which characterize the phase called Baxter-1/2.
The symbols $\langle \dots  \rangle_\textrm{F}$ and $\langle \dots  \rangle_\textrm{AF}$ indicate the thermal average of spin variables respectively in the ferromagnetic and antiferromagnetic phases, or else if $K_{4}/K_{2}>-4$, the Baxter-1/2 phase is stable since both spins $\sigma_{i}$ and $S_{i}$ are aligned and equal to 1/2.
\item[(iii)] Except for $K_{4}/K_{2}>D/K_{2}$, in the area $0<D/K_{2}<-1$, two Baxter mixed phases have been found, the first when $K_{4}/K_{2}>-0.4$, all the spins $\sigma_{i}$ and $S_{i}$ are parallel, and the second one if $K_{4}/K_{2}<-0.4$, the spins $S_{i}$ are parallel while the spins $\sigma_{i}$ are misaligned.
\end{itemize}

\section{Monte-carlo simulations}\label{sec:Monte-carlo}

The system studied here is a $L\times L$ square lattice with even values of $L$,
containing $N = L^{2}$ spins. In order to update the lattice configurations, the well-known Metropolis algorithm \cite{Bek99} is used with periodic boundary conditions. Monte-Carlo (MC) simulations are performed for $d =2$ with systems of sizes $L =10$, 16, 20, 30, 40 and 60. We use 100000 to 200000 MC steps to calculate the thermodynamic quantities after discarding 5000--50000 sweeps for thermal equilibrium. Most of the phase diagrams presented here are obtained with $L = 30$.
The physical quantities of use are the magnetizations $|M_{\alpha}|(\alpha=\sigma,S,\sigma S)$, and are estimated by:
\begin{equation}
|M_{\alpha}|=\langle|M_{\alpha}|\rangle=\frac{1}{Np}\sum_{c}\sum_{i}\alpha_{i}(c)
\qquad \text{with} \qquad \alpha=\sigma, S, \sigma S,
\end{equation}
where $i$ runs over the lattice sites, $c$ runs over the configurations obtained to update the lattice over one sweep of the $N$ spins of the lattice [one Monte-Carlo step (MCS)] counted after the system reaches thermal equilibrium, and $p$ is the number of the MCS.
In order to measure the phase boundaries, we find useful the measurement of  fluctuations (variance of the order-parameters) in $M_{\alpha}$ defined by the magnetic susceptibility:
\begin{equation}
\chi_{\alpha}=\frac{N}{k_\textrm{B}T}\left(\langle M _{\alpha}^{2}\rangle-\langle\mid M_{\alpha}\mid\rangle^{2}\right) \qquad \text{with} \qquad \alpha=\sigma, S, \sigma S,
\end{equation}
$k_\textrm{B}$ is the Boltzmann's constant.

\section{Results and discussion}\label{sec:Results}

\begin{figure}[!t]
\centerline{\includegraphics[width=0.55\textwidth]{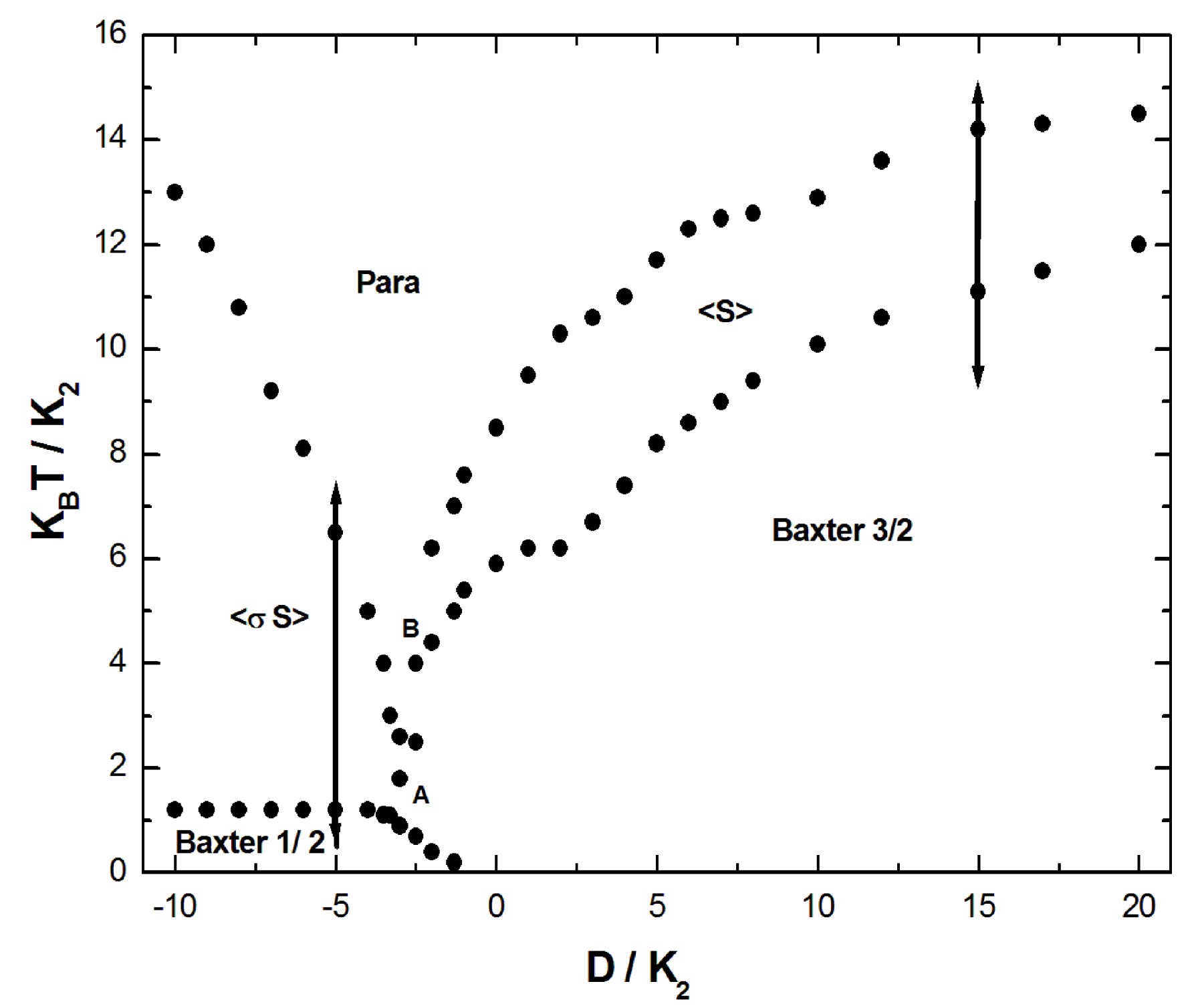}}
\caption{Phase diagram in the $(D/K_{2},T/K_{2})$ plane for $K_{4}=0.25$ from Monte Carlo simulation, data are shown with $L=30$.} \label{2}
\end{figure}

\looseness=-1The phase diagram obtained by Monte Carlo simulation is shown in figure~\ref{2} and presented in the plane $\left(k_\textrm{B}T/K_{2},D/K_{2} \right)$. We have a paramagnetic phase, where $\langle \sigma\rangle=\langle S \rangle=\langle \sigma S \rangle=0$ and two ferromagnetic (Baxter) phases, where $\langle \sigma\rangle$ and $\langle S \rangle$ are ordered ferromagnetically and also $\langle \sigma S \rangle\neq0$. The first is Baxter-1/2 and the second is the Baxter-3/2 which were not found in the earlier works \cite{Bek00,Bad99}.
These phases are separated by critical lines, multicritical points and two partially ordered phases: the $\langle \sigma S \rangle$ phase where $(\langle \sigma S \rangle\neq0$ and $\langle\sigma \rangle = \langle S \rangle = 0)$ and the $\langle S \rangle$ phase where $(\langle\sigma \rangle= \langle \sigma S\rangle = 0$, $\langle S \rangle \neq 0 $). However, the MC data are obtained from peaks in the susceptibilities \cite{Dit722} for $L=30$. The nature of the transition is determined from discontinuities (continuities) in the order parameters for first (second) order transition by MC simulations \cite{Bek99}. In our paper, the nature of the transitions is always of second order for all values.
We have located the phase boundaries by using the maximum of the susceptibility. This method has been successfully applied to other models \cite{Bek00} and has shown a good precision with the transfer matrix finite-size-scaling method \cite{Bad99}.

\begin{figure}[!b]
\centerline{\includegraphics[width=0.47\textwidth]{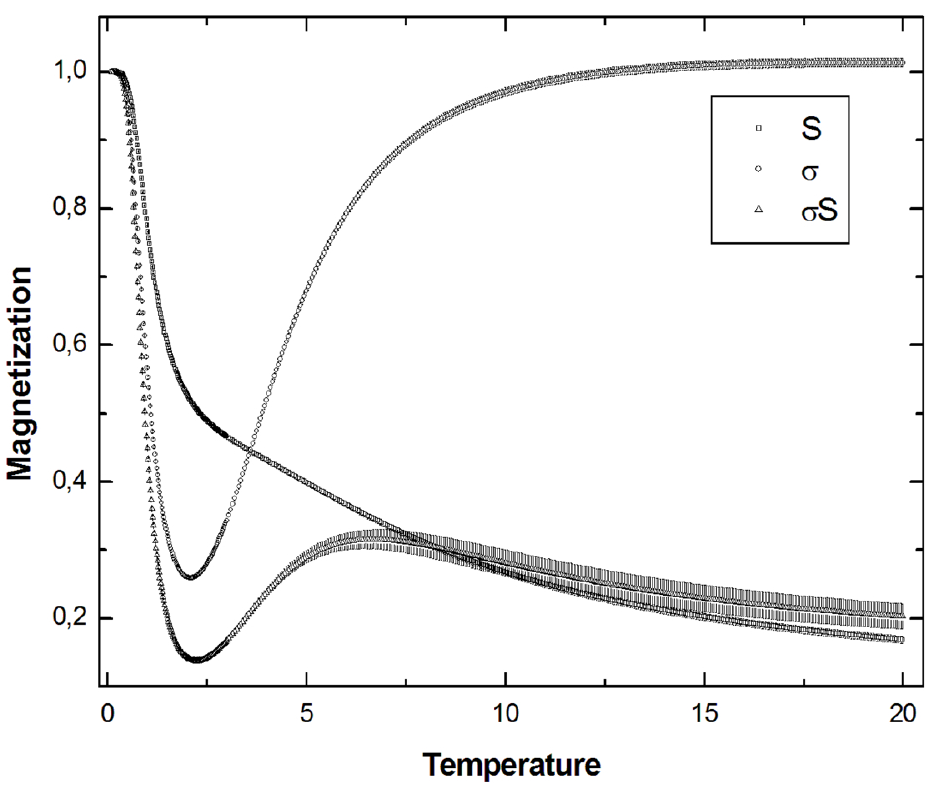}}
\caption{Plot of the three order-parameters $\langle \sigma S \rangle$, $\langle \sigma \rangle$ and $\langle S \rangle$ with error bars as function of temperature for $K_{4}=0.25$ and $D=-5$ as obtained by MC simulations for the square lattice  showing that the phase transitions are of second-order.} \label{3}
\end{figure}

\begin{figure}[!t]
\centerline{\includegraphics[width=0.51\textwidth]{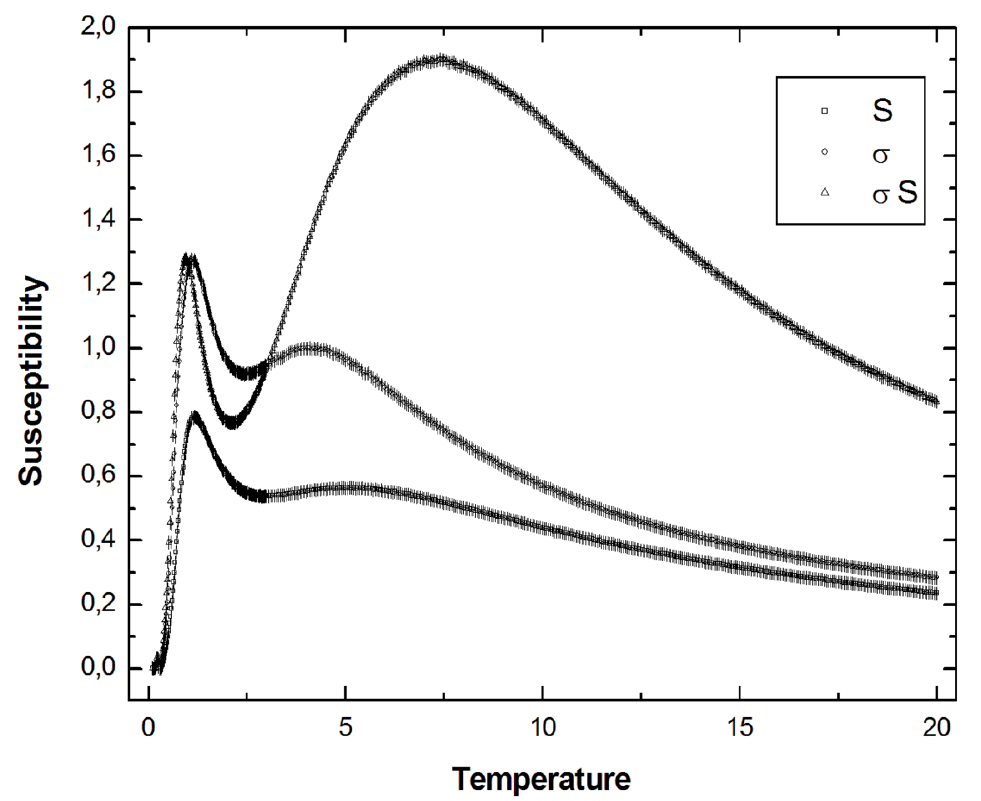}}
\caption{Plot of the susceptibility with error bars for $K_{4}=0.25$ and $D=-5$ as functions of temperature showing the existence of the phase $\langle \sigma S \rangle$, where at $T'_{1}=1.20$, $\langle \sigma S \rangle\neq0$ and $\langle\sigma \rangle = \langle S \rangle = 0$, whereas at $T''_{1}=6.48$ we have $\langle \sigma\rangle=\langle S \rangle=\langle \sigma S \rangle=0$.} \label{4}
\end{figure}

\begin{figure}[!b]
\centerline{\includegraphics[width=0.52\textwidth]{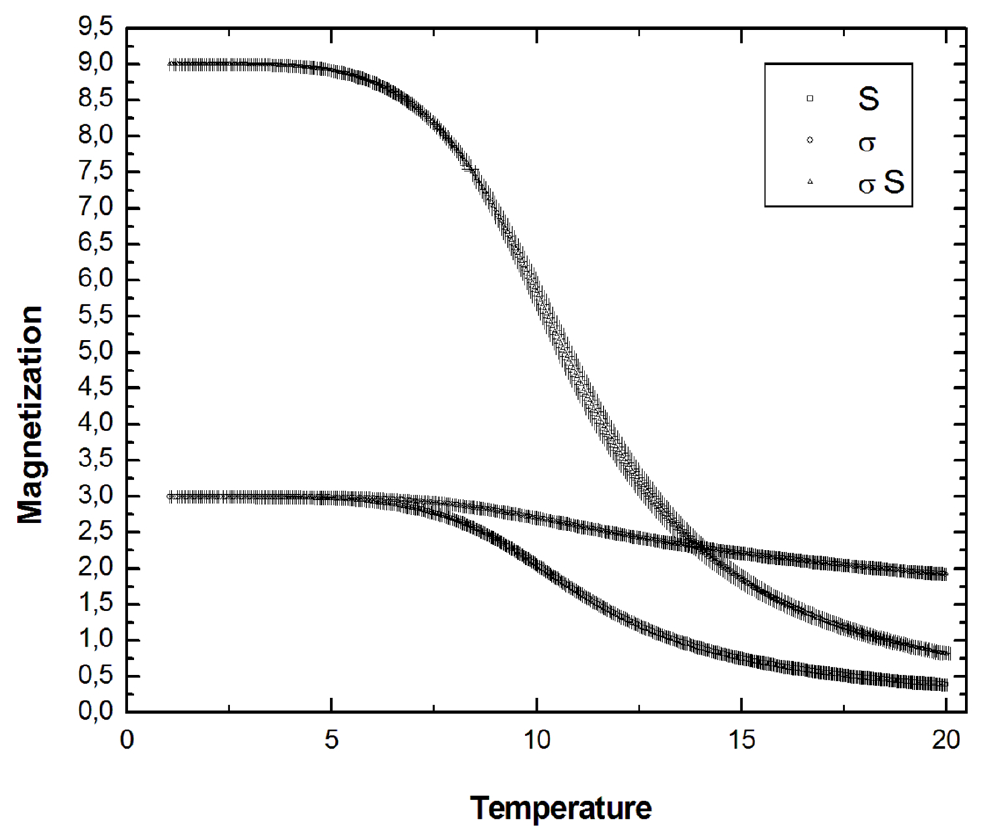}}
\caption{Plot of the three order-parameters $\langle \sigma S \rangle$, $\langle \sigma \rangle$ and $\langle S \rangle$ with error bars as function of temperature for $K_{4}=0.25$ and $D=15$ as obtained by MC simulations for the square lattice showing that the phase transitions are of second-order.} \label{5}
\end{figure}

\begin{figure}[!t]
\centerline{\includegraphics[width=0.53\textwidth]{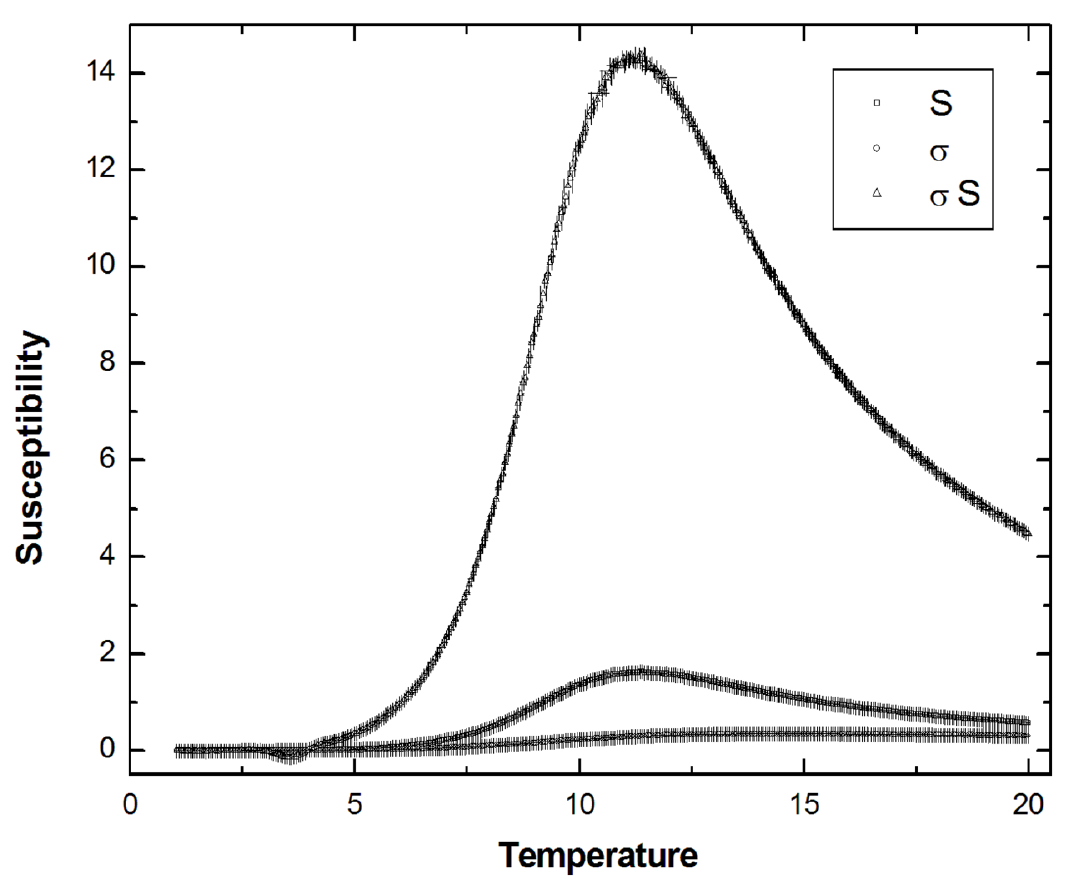}}
\caption{Plot of the susceptibility with error bars for $K_{4}=0.25$ and $D=15$ as functions of temperature showing the existence of the phase $\langle S \rangle$, where at $T'_{2}=11.10$, $\langle\sigma \rangle= \langle \sigma S\rangle = 0$, $\langle S \rangle \neq 0 $, whereas at $T''_{2}=14.21$ we have $\langle \sigma\rangle=\langle S \rangle=\langle \sigma S \rangle=0$.} \label{6}
\end{figure}

The results of figure~\ref{2} are obtained for $K_{4}=0.25$. We also see the Baxter-1/2 phase separated from the paramagnetic phase by the partially ordered phase $\langle \sigma S \rangle$ at low values of $D/K_{2}$, as seen in the figures~\ref{3}, \ref{4} and the Baxter-3/2 phase is separated from the paramagnetic phase by the new partially ordered phase $\langle S \rangle$ at high values of $D/K_{2}$, as seen in the figures~\ref{5}, \ref{6}. This phase does not exist either in the spin-1/2 \cite{Bek99} or in the spin-1 \cite{Bad99} Ashkin-Teller model but is viewed in the mixed (ATM) \cite{Bek00}.
The meeting of all critical lines is located at the  multicritical points $A(-2.98\pm0.01;0.77\pm0.01)$ and $B(-2.54\pm0.01;3.97\pm0.01)$. The two phases  $\langle \sigma S \rangle$ and  $\langle S \rangle$ separate the disordered phase from the ferromagnetic Baxter phases by second order transition lines.

For error bars, we have made 300000 MCS and discarding 50000 and made measurements every 100 MCS, we plotted errors to the magnetizations and susceptibilities, as seen in the figures~\ref{3}--\ref{6} which present plot of the three order-parameters $\sigma$, $S$ and $\sigma S$ as function of temperature for $K_4=0.25$ and $D=-5$ (and $D=15$) as obtained by MC simulations showing that the error bars are very small, the simulation was too long, it took a lot of computing time.

\begin{figure}[!b]
\centerline{\includegraphics[width=0.5\textwidth]{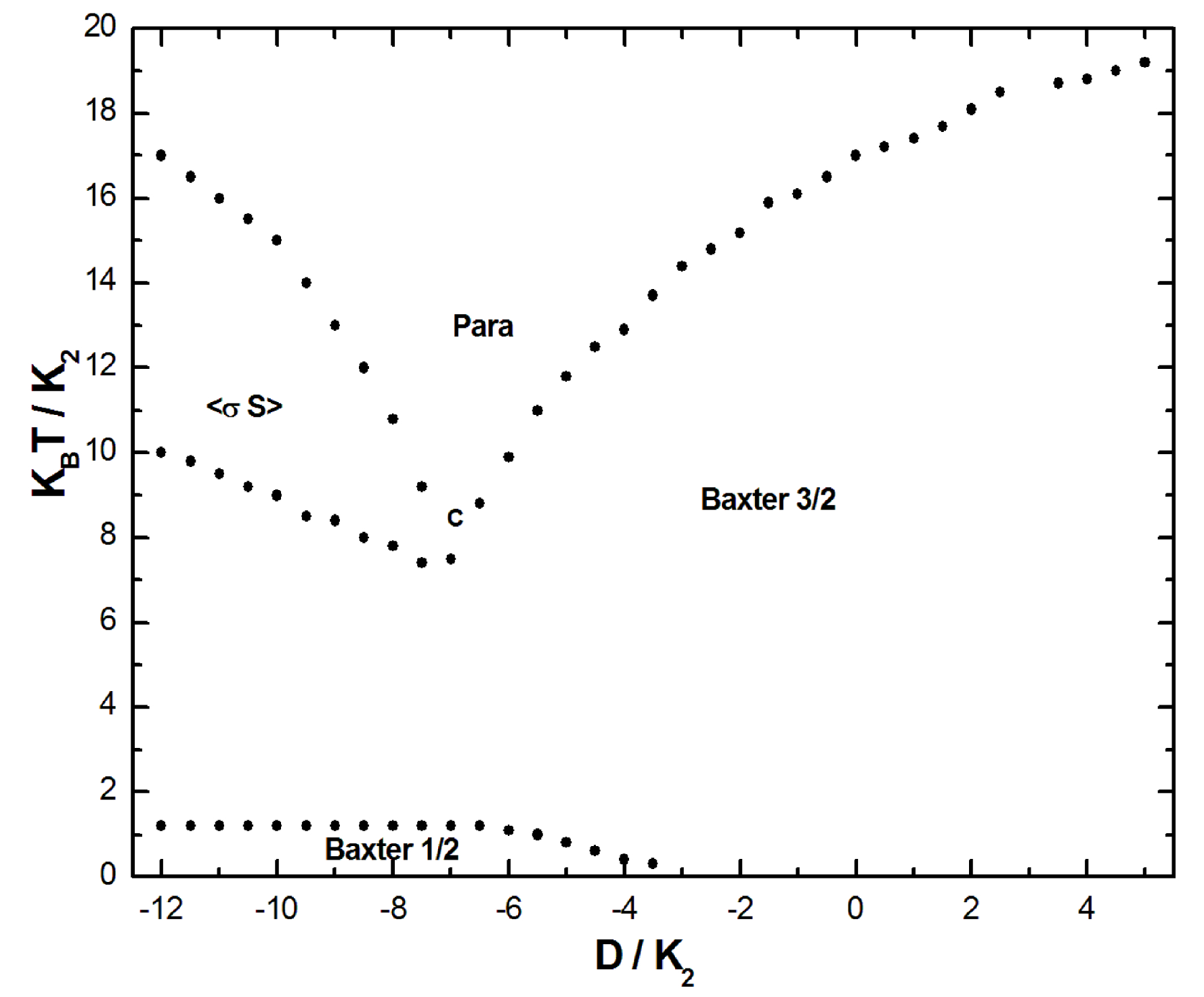}}
\caption{Phase diagram in the $(D/K_{2},T/K_{2})$ plane for $K_{4}=2$ from Monte Carlo simulation, data are shown with $L=30$.} \label{7}
\end{figure}

\begin{figure}[!t]
\centerline{\includegraphics[width=0.53\textwidth]{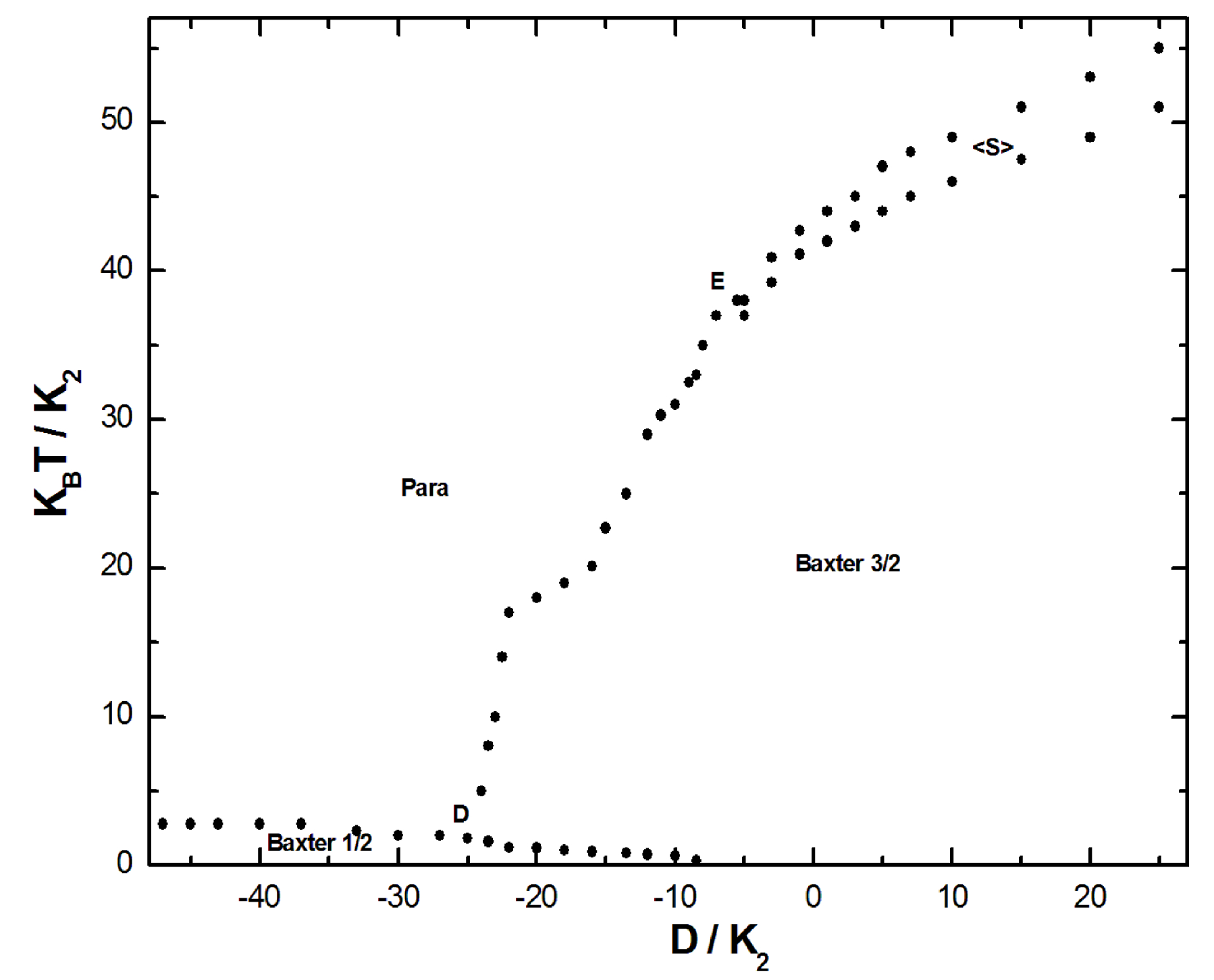}}
\caption{Phase diagram in the $(D/K_{2},T/K_{2})$ plane for $K_{4}=6$ from Monte Carlo simulation, data are shown with $L=30$.} \label{8}
\end{figure}

By increasing the four body interaction, $K_{4}/K_{2}$, the two Baxter phases remain in the phase diagrams whereas depending on the strength of $K_{4}$, as shown in figure~\ref{7} for $K_{4}=2$, the partially ordered phase $\langle \sigma S \rangle$ shrinks and the other one $\langle S \rangle$ disappears. But  for strong $K_{4}$, $K_{4}=6$, the partially ordered phase $\langle \sigma S \rangle$ disappears and the other partially phase $\langle S \rangle$ is recovered as shown in figure~\ref{8}. Continuous lines are linked by multicritical points of higher order.

\section{Conclusion}\label{sec:Conclusion}

In this paper, by using MC simulations we have shown that the isotropic ferromagnetic Ashkin-Teller model  presents a new partially ordered phase $\langle S \rangle$, which is very clear at high temperatures (also found infinitesimal in the Monte Carlo mixed ATM \cite{Bek00}), and other phases like Baxter-3/2 where all spins have the magnitude of 3/2. In the parameter space ($K_{4} /K_{2} $, $D/K_{2}$ and $T/K_{2})$, the phase diagrams present rich varieties of phase transitions with surfaces of second order phase transitions, bounded by lines of  multicritical points.

In conclusion, the study was carried out on the model of Ashkin-Teller spin-3/2. It has revealed the complexity of the model with very rich structures and gives a better understanding of the properties of condensed matter, especially magnetic properties of the systems consisting of many atom molecules holders.
It is on the basis of this study that we plan to study the magnetic properties of the Ashkin-Teller model with mixed spins on different types of lattices. The presence of many atoms with different magnetic moments on the  same site can reveal some interesting properties. The model will also be analyzed in three dimensions including the crystal fields and long range interactions.

\clearpage

\ukrainianpart

\title[]%
{Чисельне вивчення спін-3/2 моделі Ашкіна-Теллера}

\author[]%
{Р. Будефля\refaddr{label1}, С. Бекеші\refaddr{label1}, Ф. Гонтінфінде\refaddr{label2}
}
\addresses{
\addr{label1} Лабораторія теоретичної фізики, B.P. 230, Університет Абу Бакр Белькаїд, Тлемсен 13000, Алжир
\addr{label2} Відділення фізики (FAST) та інститут математики і фізичних наук  (IMSP), Університет Абомей-Калаві, Бенін}

\makeukrtitle

\begin{abstract}
Вивчення моделі Ашкіна-Теллера спін-3/2 на гіперкубічній ґратці здійснюється методом Монте Карло. Продемонстровано фазові діаграми і обговорено простір фізичних параметрів. Виявлено багатство фізичних властивостей, а саме, перехід другого роду і мультикритичні точки. Фазові діаграми отримані шляхом зміни сили, що описує чотириспінову взаємодію та одноіонний потенціал.
Дана модель демонструє нову високотемпературну частково впорядковану фазу, що називається  $\langle S\rangle$, і новий основний 3/2 стан Бакстера, якого нема в  моделях Ашкіна-Теллера спін-1/2 і  spin-1.
\keywords моделювання, Ашкін-Теллер, спін-3/2, Монте-Карло, фазова діаграма, Бакстер
\end{abstract}
\end{document}